\documentclass[aps,draft,reprint]{revtex4-1}
 
\usepackage{latexsym,hyperref,graphicx}
 \hypersetup{colorlinks,citecolor= nicegreen,linkcolor= nicered}
\usepackage{color}
\definecolor{nicered}{rgb}{0.7,0.1,0.1}
\definecolor{nicegreen}{rgb}{0.1,0.5,0.1}

 \newcommand{\beq}{\begin{equation}}
\newcommand{\eeq}{\end{equation}}

\newcommand{\beqa}{\begin{eqnarray}}
\newcommand{\eeqa}{\end{eqnarray}}

\usepackage{amsmath}

\begin{document}

\title{A note on Spontaneous Symmetry Breaking in flocks of birds}

\author{Alejandra Melfo} 
 
\affiliation{Centro de F\'isica Fundamental, Universidad de Los Andes,
M\'erida, Venezuela}

\date{\today}

\begin{abstract}
It is suggested that  the observed scale-free correlations of speed fluctuations in flocks of birds are a consequence of the spontaneous breakdown of translational symmetry to a discrete group, and not an indication that the system is near a critical point in phase space. The observed long-range correlation length could then be attributed to the presence of a phonon mode in the flock.

\end{abstract}

\maketitle

It has been  pointed out   that some biological systems tend to be situated at critical points in phase space, the relevant parameters fine-tuned to produce a collective effect. Such is argued to be the case  in networks of neurons, ensembles of protein-coding DNA sequences, and flocks of birds \cite{morabialek}. In the latter case, the argument for criticality is based on analysis of the bird's collective  motion using techniques borrowed from  statistical physics, following the seminal work in this direction by Vicsek and collaborators \cite{vicsek} and by Toner and Tu \cite{toneryu}, and on recent observations of starling ({\it Sturnum vulgaris}) flocks by the STARFLAG group in Rome \cite{cabibbo,starlians,bialek14}. 

The key observations are presented in \cite{starlians}. Data from flocks of starlings of different sizes, ranging from  around 100 to 4000 individuals, was obtained using stereometric digital photogrammetry and computer vision techniques. For each of the 24 flocking events observed, a complete reconstruction of the individual positions   and velocities  of birds in the flock was performed.
 By subtracting the instantaneous velocity of the center of mass, the fluctuations in velocity with respect to the mean are obtained for each bird. The   two-point correlation functions of the velocity and speed fluctuations are then found, and correlation distances $\xi_{V}  $ for orientation and  $\xi_S$ for speed are  calculated as the distance for which the respective correlation function vanishes.

On the other hand, regardless of their specific nature, the bird's interactions have been shown to be short range, depending on the closest neighbors. The correlation functions for speed and orientation fluctuations are {\it a priori} expected to change sign many times across the flock, providing a correlation distance that could in principle be of the order of the interaction distance. However, both correlations were found to be scale invariant, with  $\xi_{V}  $  and  $\xi_S$   of the order of the flock's size. In statistical physics systems, such an ``infinite" correlation length is what one would expect   near a critical point in phase space, in the sense of Landau-Ginzburg theory. For birds, behaving collectively as if near a critical point would entail a precise fine-tuning of the parameters governing their interactions.

There is a caveat, however, already pointed out by Toner and Tu \cite{toneryu} and duly emphasized by the authors of \cite{starlians}: infinite correlation length can also be present away from critical points, if the system has undergone the Spontaneous Symmetry Breaking (SSB) of a continuous symmetry. In that case, the infinite correlation length can be seen as a manifestation of a massless mode propagating in the system, the Nambu-Goldstone boson. In quantum field theory, the Nambu-Goldstone boson is a massless particle connecting the equivalent vacua. In a flock of birds, it is just the fact that it costs no energy for an individual bird to perform a certain change in their flight. Thus, ever since the first studies on the subject, it has been suggested that a group of birds that spontaneously align their velocities breaks the rotational symmetry spontaneously, and the resulting Nambu-Goldstone mode should ensure that the correlation lengths of the orientations are of order of the system's size. 

This is exactly what has been observed: birds spontaneously aligning their velocities break the rotational symmetry just as spins in a ferromagnetic material, one of the first  examples of SSB studied, do. Thus the fact that  $\xi_{V}  $  is large does not come as a surprise. It is the fact that $\xi_S$, the correlation length of speed fluctuations, is also large, what leads the authors to their conclusion that the flock is critical: the SSB of a rotational symmetry does not produce correlations in the relevant vector's magnitude. In other words, the Nambu-Goldstone mode is related to the bird's not having any preference for alignment in a particular direction (which expresses the rotational symmetry of the Hamiltonian), and thus requiring no difference in energy to align in one direction respect to the other (which expresses the massless of the Nambu-Goldstone mode).  
But rotational symmetry is not related to the speed. In  \cite{bialek14} a model of maximum entropy for flocks is presented, and the authors use as input the parameters obtained from observations in \cite{starlians} to demonstrate the critical behavior. It is noted that: {\it ``Not just the fluctuations in flight direction, but also the fluctuations in flight speed are correlated over long distances. Now there are no Goldstone modes, because choosing a speed does not correspond to breaking any plausible symmetry of the system"}.  Thus the infinite correlation lengths, the authors conclude, must be a sign of criticality. 

Collective motion is however  widespread  in biological systems (see {\it e.g.} \cite{couzin} and references therein), and the proposal that flocking behavior can happen only at critical points in parameter space seems to be at at odds with simulations, where it frequently results from a wide choice of parameters. Admittedly, it is not possible to change the parameters governing real bird's interactions to check if the statistical mechanics treatment holds and they are indeed fine-tuned. This can only be done in models such as those of Hemelrijk and collaborators \cite{hemelrijk1, hemelrijk2},  where agents that obey simple rules can reproduce the observations on the Rome starlings.  Most simulations of flocking belong to the class of Vicsek models \cite{vicsek}, in which agents align in the direction of their closest neighbors.   Recently, a series of alternatives have appeared (for a review, see \cite{vicsekreport}), many of them belonging to the category of position-based models. One simple example is provided in \cite{huepeII}: an active-elastic model where birds interact with a restoring force, that changes sign as their relative distance falls below a threshold, thus forcing the agents to keep a fixed distance with one another. This spring-like interactions result in different kinds of collective motion for different initial conditions, and require no fine tuning of the parameters. 

In \cite{positionbased}, such a position-based model is used to simulate a flock of birds and the two-point correlation functions for orientation and speed are calculated. The results of \cite{starlians} are recovered, namely  the correlation lengths $\xi_V$ and $\xi_S$ are of the size of the flock, without need of fine tuning. The authors argue that criticality is not necessary, and that the results of  \cite{starlians,bialek14}   {\it  ``could be a consequence of imposing a model that can only display long-range speed correlations near criticality on a system that develops such correlations for a different reason".} 

We point out that this reason could   be the SSB of the translational symmetry. Namely, flocking exhibit three features commonly referred to as ``the three A's": alignment, attraction and avoidance. When the agents align, they break spontaneously the rotational symmetry; when they  attract or avoid each other, thus keeping a distance with their neighbors (irrespectively of whether if interactions are spatial or topological, \cite{cabibbo}), they break spontaneously the continuous translational invariance to a discrete group. The system is thus analogous to a lattice, and the Nambu-Goldstone mode is  the  phonon. This mode will result in scale-free correlations in fluctuations of the instantaneous relative positions, and therefore in the speed.

One could worry that a proof of the Goldstone theorem  generally assumes translational invariance. Phonons are however standard in condensed matter physics, and their behavior as Goldstone modes is studied for example in \cite{leutweyler}. The interested reader can find a proof  of the validity of the Goldstone theorem when translational symmetry is broken to a discrete group in  \cite{watanabe}.  

If long-range correlations arise from SSB of the translation group, they should appear in other examples of collective motion, both real and modeled, as by the very definition of collective motion the relative velocity of the individual particles should be zero, implying a constant distance. It would be interesting to measure correlations in collective motion of other species. It would also be interesting to find  a further consequence of SSB, to ensure that the discussion is more than semantic. In the case of rotational symmetry, topological defects such as vortices (milling behavior in fish) could arise. The translation group is however not compact, and the usual theorems predicting topological defect formation on general grounds do not apply.

Admittedly, although the Goldstone theorem is based on very general considerations, requiring nothing more than a continuous symmetry of the Hamiltonian that is not manifest in the ground state, its validity in a system of entities that have a finite size, that come in relatively small numbers,  forming a non-isolated system with a finite extension, can be doubtful. The same can be said for any treatment of a flock of birds as a statistical mechanics system. The important point is that if SSB of rotational symmetry is accepted as responsible for the large correlation lengths in orientation fluctuations observed in real flocks, then SSB of translational symmetry should be equally considered. Doing so could contribute to solve the controversy between the argument for criticality given in \cite{starlians,bialek14} and results as the one presented in \cite{positionbased}.

 \subsection*{Acknowledgements}
I am indebted to Paulien Hogeweg for her suggestion to look at this problem, discussions and encouragement.  I am also grateful to Goran Senjanovi\'c for discussions, and to the Theoretical Biology and Bioinformatics group at the University of Utrecht for their hospitality during this work.

\end{document}